\documentclass[sigconf, nonacm]{acmart}

\usepackage[capitalise, noabbrev]{cleveref}
\usepackage{tikz}
\usetikzlibrary{backgrounds,positioning,calc}
\definecolor{extcolor}{RGB}{210,70,55}    
\definecolor{intcolor}{RGB}{45,85,175}    
\usepackage{subcaption}

\newcommand\graphalg{\textsc{GraphAlg}}
\newcommand\javascript{JavaScript}

\newcommand*\numcircledtikz[1]{\tikz[baseline=(char.base)]{
            \node[shape=circle,draw,inner sep=1.2pt] (char) {#1};}} 

\newcommand*\step[1]{\noindent \textbf{Step #1.}}

\newcommand\vldbvolume{19}

\newcommand\vldbavailabilityurl{https://wildarch.dev/graphalg}

\begin{document}
\title{GraphAlg Playground: An Online Platform for Learning and Experimenting with the GraphAlg Language}

\author{Daan de Graaf}
\orcid{0009-0000-0322-1149}
\email{d.j.a.d.graaf@tue.nl}
\affiliation{%
  \institution{Eindhoven University of Technology}
  \city{Eindhoven}
  \country{Netherlands}
}

\author{Robert Brijder}
\email{r.brijder@tue.nl}
\affiliation{%
  \institution{Eindhoven University of Technology}
  \city{Eindhoven}
  \country{Netherlands}
}

\author{Soham Chakraborty}
\email{s.s.chakraborty@tudelft.nl}
\affiliation{%
  \institution{Delft University of Technology}
  \city{Delft}
  \country{Netherlands}
}

\author{George Fletcher}
\email{g.h.l.fletcher@tue.nl}
\affiliation{%
  \institution{Eindhoven University of Technology}
  \city{Eindhoven}
  \country{Netherlands}
}

\author{Bram van de Wall}
\email{a.a.g.v.d.wall@tue.nl}
\affiliation{%
  \institution{Eindhoven University of Technology}
  \city{Eindhoven}
  \country{Netherlands}
}

\author{Nikolay Yakovets}
\email{hush@tue.nl}
\affiliation{%
  \institution{Eindhoven University of Technology}
  \city{Eindhoven}
  \country{Netherlands}
}

\begin{abstract}
  The \graphalg{} language for graph algorithms enables native support for user-defined graph analytics workloads in databases.
  In this demonstration, we present a web-based playground for writing and executing \graphalg{} programs in the web browser, including an interactive tutorial explaining its key concepts.
  The playground runs inside the user's web browser without any installation, and is freely available under a permissive license as a reusable library.
  We present two demonstration scenarios of the publicly available playground website, showing how new users can learn to program in \graphalg{} using the tutorial, while expert users can use the playground to prototype and validate their algorithms.


\end{abstract}

\maketitle

\pagestyle{plain}
\begingroup\small\noindent\raggedright
Accepted at the VLDB 2026 Demonstration Track. To appear in the Proceedings of the VLDB Endowment, Vol.~\vldbvolume{} (VLDB 2026). Artifacts available at \url{\vldbavailabilityurl}.
\endgroup

\section{Introduction}
Consider a data scientist analyzing citation networks to identify influential papers using PageRank, or a fraud analyst running community detection on transaction graphs.
Despite graph databases being the natural home for such data, these users cannot express these fundamental algorithms in Cypher~\cite{francis_cypher_2018} or GQL~\cite{iso_information_2024}.
Instead, they must export gigabytes of data and wrestle with format conversions.
This ``data wrangling'' wastes engineering effort and risks correctness when copies diverge from the source of truth.

\begin{figure}[t]
  \centering
  \includegraphics[width=.6\linewidth]{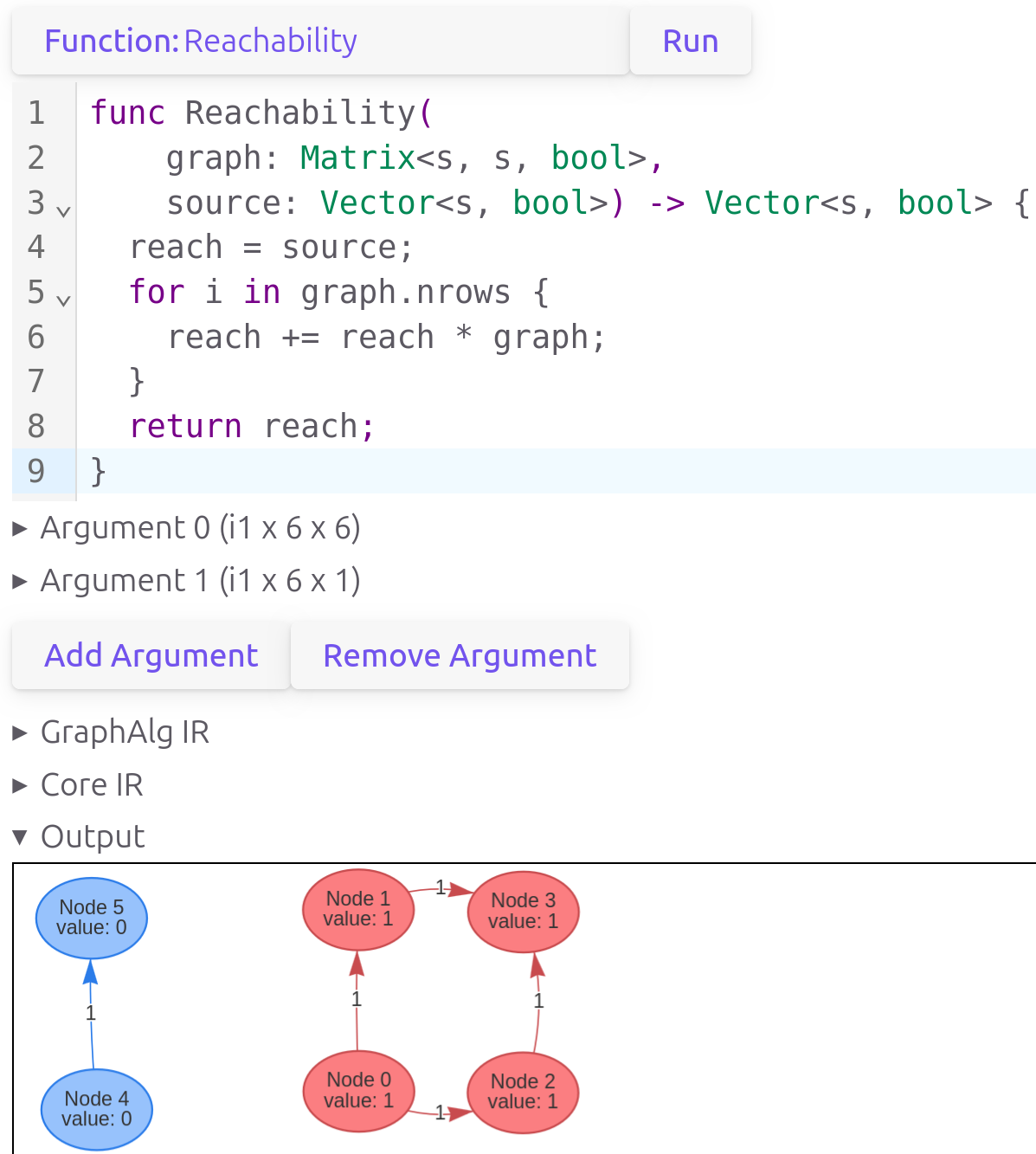}
  \caption{A \graphalg{} program in the playground.}
  \label{fig:reach}
  \Description{
    A reachability algorithm written in the \graphalg{} playground code editor.
    The output of the algorithm is shown below it as a network visualization.}
\end{figure}

\begin{figure}[t]
  \centering
  \includegraphics[width=.8\linewidth]{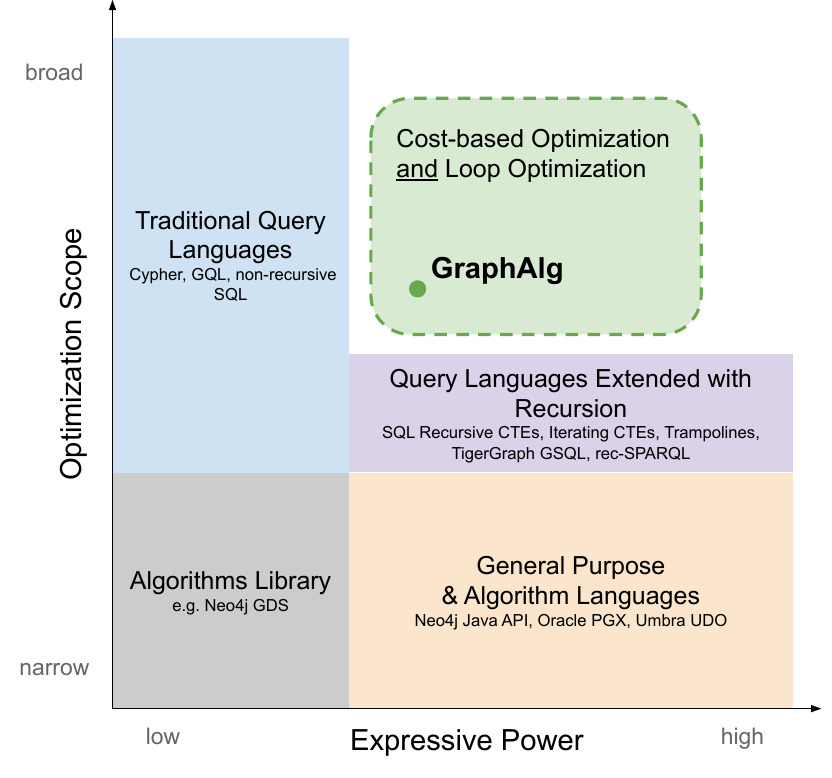}
  \caption{
    Existing approaches to in-database graph analytics along two dimensions:
    the complexity of algorithms they support (horizontal) and how automatically they choose efficient execution strategies (vertical).
    \graphalg{} aims toward the upper-right corner, carefully balancing expressive power with a high degree of automated optimization.}
  \label{fig:landscape}
  \Description{A plot positioning existing approaches to in-database graph analytics along two axes: expressive power (horizontal) and degree of automated optimization (vertical). GraphAlg is placed in the upper-right corner.}
\end{figure}

Existing approaches to graph analytics support in databases suffer from various problems.
Algorithm packages such as the Neo4j Graph Data Science Library~\cite{neo4j_inc_graph_nodate} offer only fixed implementations that rarely match exact requirements (e.g., Neo4j's PageRank lacks sink redistribution).
Other systems integrate general-purpose programming languages or separate algorithm languages~\cite{sichert_user-defined_2022,neo4j_inc_graph_nodate} for maximum flexibility, but this impedes cost-based optimization inside algorithms.
Another popular approach is to extend existing query languages with recursion~\cite{hirn_fix_2023,gupta_procedural_2021,ma_g-sql_2016,hogan_-database_2020}.
This approach inherits the cost-based optimizations from the original query languages, but leaves high-level optimization across loop iterations to the programmer.
Finally, the existing systems that both support loops \emph{and} apply comprehensive optimization~\cite{shaikhha_optimizing_2024,hutchison_laradb_2017,shkapsky_optimizing_2015} are standalone solutions that do not integrate with graph data management systems.
\cref{fig:landscape} provides a visual overview of the database-native approaches according to their expressive power and the scope of automated optimization.

\graphalg{}~\cite{de_graaf_algorithm_2026} is a domain-specific language for graph algorithms, based on linear algebra, that addresses this gap.
\graphalg{} satisfies four critical requirements:
(1)~\textbf{Expressive}: users implement arbitrary algorithms by composing matrix operations;
(2)~\textbf{Designed for graph algorithms}: programs require 2--10$\times$ less code than SQL or Java;
(3)~\textbf{Fully integrated}: algorithms embed directly in queries and execute on the native graph storage;
(4)~\textbf{Optimizable}: the \graphalg{} compiler performs loop optimization, after which programs compile to query plans to leverage existing database optimizations.
We integrated \graphalg{} into the AvantGraph~\cite{van_leeuwen_avantgraph_2022} graph data management system, where it outperforms DuckDB and Neo4j on PageRank, single-source shortest paths, and connected components benchmarks~\cite{de_graaf_algorithm_2026}.
We implemented a diverse set of centrality, community detection and path finding algorithms in \graphalg{} widely considered as representative for graph analysis in general~\cite{iosup_ldbc_2016}.

To streamline the process of learning the language,
we propose the \graphalg{} \emph{playground}: an online platform for learning and experimenting with \graphalg{}.
We adopt a fully client-side implementation based on WebAssembly~\cite{haas_bringing_2017} that eliminates network overhead, enabling real-time linting and error diagnostics.
\cref{tab:other-playgrounds} compares the \graphalg{} playground to other online interfaces for programming languages and database systems where users can write and execute programs from their browser without installing any software~\cite{google_inc_go_nodate,the_rust_developers_rust_nodate,the_umbra_developers_umbra_nodate,the_duckdb_developers_duckdb_nodate}.
The playground supports the full \graphalg{} language, so any valid \graphalg{} program runs in it; its only limitation is an unoptimized WebAssembly runtime, making it suited to learning and small-scale experimentation rather than the production workloads AvantGraph handles.

\begin{table}
  \footnotesize
  \centering
  \caption{
    Features of online playgrounds for several programming languages and database systems.
  }
  \begin{tabular}{|l|c|c|c|c|c|}
    \hline
                                  & \textbf{Go}                                   & \textbf{Rust}                                    & \textbf{Umbra} & \textbf{DuckDB} & \textbf{\graphalg{}} \\
    \hline
    \textbf{Execution}            & \multicolumn{3}{c|}{Server-side}              & \multicolumn{2}{c|}{\textbf{Client-side (WASM)}}                                                           \\
    \hline
    \textbf{Syntax Highlighting}  & No                                            & \multicolumn{4}{c|}{\textbf{Yes (Client-side)}}                                                            \\
    \hline
    \textbf{Compiler Diagnostics} & \multicolumn{4}{c|}{When clicking run button} & \textbf{Real-time}                                                                                         \\
    \hline
    \textbf{Error Highlighting}   & \multicolumn{4}{c|}{No}                       & \textbf{Yes}                                                                                               \\
    \hline
    \textbf{Result Rendering}     & \multicolumn{4}{c|}{Text (stdout)}            & \textbf{Graph viz.}                                                                                        \\
    \hline
  \end{tabular}
  \label{tab:other-playgrounds}
  \Description{
    A table comparing different online playgrounds on the following features.
    Execution: Go, Rust and Umbra execute clients on a remote server, whereas DuckDB and GraphAlg execute client-side using WebAssembly.
    Syntax Highlighting: Go does not offer syntax highlighting, all others do and run it client-side to highlight as the user types.
    Compiler Diagnostics: \graphalg{} provides compiler diagnostics in real time, while other systems only provide feedback after the users clicks the run button.
    Error Highlighting: Only \graphalg{} underlines the source location where the error originates.
    Result Rendering: Where the other playground output plain text (standard output), \graphalg{} renders a network visualization.
  }
\end{table}

The \graphalg{} playground includes an \emph{interactive} tutorial that explains key language concepts, starting from basic syntax and gradually increasing complexity up to complete, practical algorithms such as PageRank.
Every part of the tutorial is accompanied by interactive \graphalg{} code snippets that can be modified and executed directly from the user's web browser.

The \graphalg{} playground also serves experienced programmers already familiar with the language.
The same editor with syntax highlighting and real-time linting provides a platform for developing new algorithms.
Programmers can upload input data (e.g., a small test graph), run the algorithm, and visually inspect output matrices rendered as graphs.
This rapid feedback loop accelerates algorithm prototyping without requiring a full database setup.

The entire playground, including the compiler, runtime, and tutorial, is freely available under a permissive open-source license~\cite{de_graaf_wildarchgraphalg_2026}.

Our demonstration consists of two scenarios:
(1) Using the playground as an experienced \graphalg{} programmer to modify and test an algorithm, and
(2) Working through the tutorial as a developer new to \graphalg{}.

\section{GraphAlg in a Nutshell}
\graphalg{} is a domain-specific language for graph algorithms based on linear algebra.
Matrix operations naturally express graph computation: multiplying a vector of ``reached'' vertices by the adjacency matrix yields the vertices reachable in one hop.
This correspondence facilitates concise definitions of algorithms.

The language provides high-level operations (matrix multiplication, aggregation, element-wise application) with an imperative style similar to Python.
For example, the reachability algorithm shown in \cref{fig:reach} is just a loop that repeatedly multiplies the frontier vector by the adjacency matrix and accumulates results.
A powerful feature is support for \emph{semirings}~\cite{kepner_graph_2011}, which generalize arithmetic operations: using the tropical semiring (with $\min$ for addition and $+$ for multiplication) transforms a reachability algorithm into single-source shortest paths with minimal code changes.
This expressiveness allows \graphalg{} to implement algorithms including PageRank, connected components, community detection, and breadth-first search.

Programs compile to relational algebra via a formal core derived from MATLANG~\cite{brijder_expressive_2019}, enabling integration into existing query pipelines.
In AvantGraph, \graphalg{} programs embed directly in Cypher queries and are optimized together with the surrounding query, enabling cross-boundary optimizations impossible in systems with separate algorithm pipelines.

\begin{figure}[t]
  \centering
  \includegraphics[width=.82\linewidth]{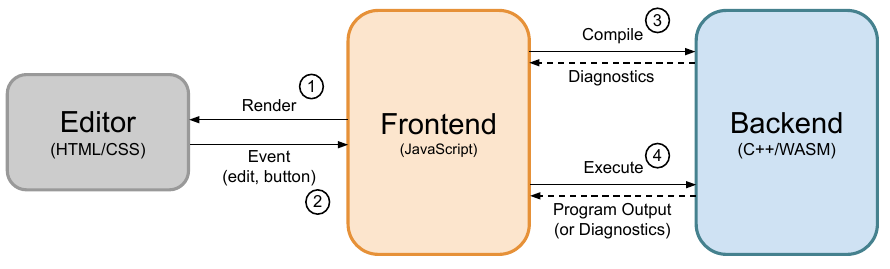}
  \caption{Overview of the \graphalg{} Playground Architecture.}
  \label{fig:arch}
  \Description{
    A Frontend component (in JavaScript) renders an Editor (HTML/CSS) and receives user input from that Editor.
    The Frontend interfaces with a Backend component (in C++/WebAssembly) to compile and execute graphalg programs.
  }
\end{figure}

\section{Playground Architecture}

The \graphalg{} playground architecture (\cref{fig:arch}) consists of two components:
a \emph{Frontend} (\javascript{}) that renders interactive code editors and handles user input, and a \emph{Backend} (C++/WebAssembly) that parses, compiles, and executes \graphalg{} programs.

Written in \javascript{}, the Frontend is included as a script in an HTML document.
Upon loading, it scans the document for \graphalg{} code markers and \numcircledtikz{1} renders interactive editors in their place.
This approach allows the tutorial to be written in Markdown, with code snippets automatically becoming interactive editors.
The Frontend also downloads and initializes the Backend.

The Backend is compiled to WebAssembly so it runs entirely client-side, avoiding the need for server infrastructure.
This eliminates hosting costs and scalability concerns, and keeps user code fully private since it never leaves the browser.
The Backend uses the same open-source \graphalg{} compiler library~\cite{de_graaf_wildarchgraphalg_2026} as AvantGraph, ensuring programs validated in the playground work identically in production.
We illustrate the components working together through two key features.

\begin{figure}[h]
  \centering
  \includegraphics[width=.68\linewidth]{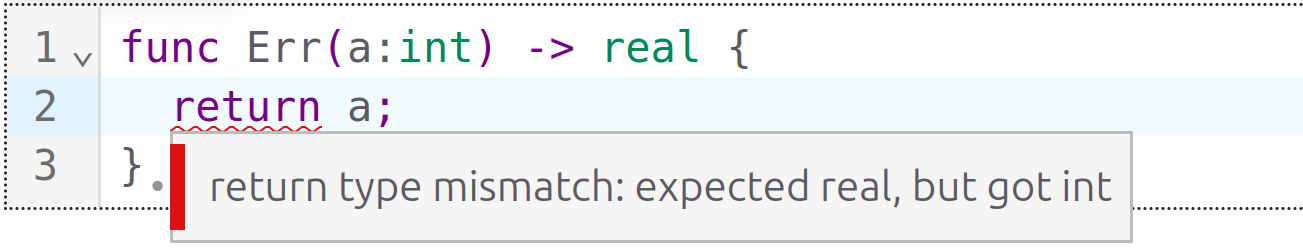}
  \caption{An error diagnostic in the code editor, generated by type checking in the \graphalg{} compiler.}
  \label{fig:diagnostics}
  \Description{An error message displayed as a tooltip in the code editor. The message is 'return type mismatch: expected real, but got int'}
\end{figure}

\paragraph{Real-time Compiler Feedback}
When the user edits code \numcircledtikz{2}, the Frontend requests the Backend to \numcircledtikz{3} compile the program.
If parsing or type checking fails, the Backend returns error diagnostics with source locations.
The Frontend \numcircledtikz{1} underlines errors in the editor; hovering over the underlined text displays the error message in a tooltip (\cref{fig:diagnostics}).
Because compilation happens locally via WebAssembly, feedback is instantaneous.
The \graphalg{} type system catches errors such as dimension mismatches at compile time; remaining runtime errors are reported with source locations.

\begin{figure}[h]
  \centering
  \includegraphics[width=.68\linewidth]{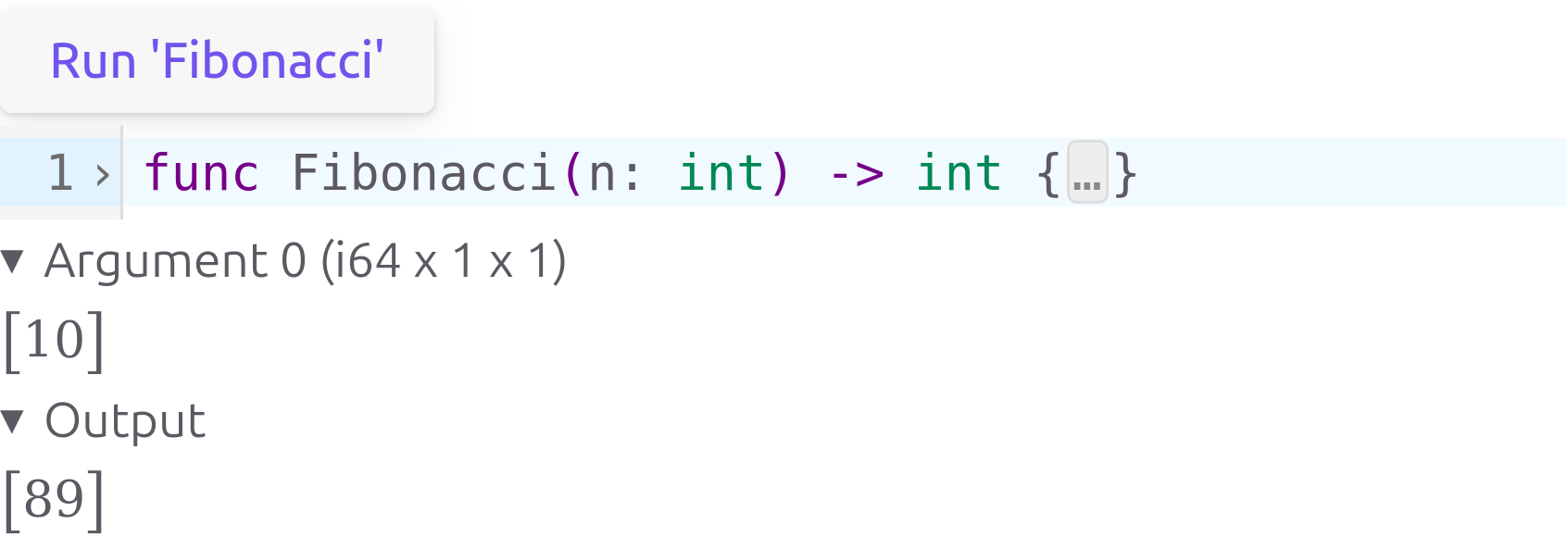}
  \caption{Code editor with rendered argument and output.}
  \label{fig:output}
  \Description{A code editor with a fibonacci sequence program. The input argument is '10' and the output is '89'.}
\end{figure}

\paragraph{Algorithm Execution in the Browser}
When the user clicks the \emph{Run} button \numcircledtikz{2}, the Frontend collects input arguments and requests the Backend to \numcircledtikz{4} execute the program.
The Backend compiles and runs the program, returning output to the Frontend.
The Frontend \numcircledtikz{1} renders results below the code (\cref{fig:output}).
Matrix outputs can be visualized as a graph, making algorithm behavior easy to inspect, and users can upload their own graph to test algorithms on custom inputs.

\newcommand\expuser{Alex}
\newcommand\newuser{Beau}

\section{Demonstration}
Our demonstration comprises two scenarios highlighting different use cases for the \graphalg{} playground.
The first scenario shows developing and testing a \graphalg{} algorithm in the playground, then deploying it on a large graph in a production-like setup.

In the second (optional) scenario, attendees work through exercises in the \graphalg{} tutorial, which is publicly available and fully web-based, so participants can also continue on their own device later.

\subsection{Modifying, Testing and Deploying an Algorithm using the Playground}
\expuser{} maintains a scientific knowledge graph with tens of millions of citations using the AvantGraph~\cite{van_leeuwen_avantgraph_2022} graph data management system.
To find high-impact publications in the graph, they apply PageRank to the citation graph, implemented as a \graphalg{} program embedded in a Cypher query.

\expuser{} wants to extend their PageRank implementation to redistribute scores from sink nodes (a common requirement that built-in algorithm libraries do not satisfy~\cite{de_graaf_algorithm_2026}).
They take the following steps to modify the existing program, test the behavior of the modified program, and finally deploy it on the large graph.

\step{1} \expuser{} opens their browser and navigates to the playground at \href{https://wildarch.dev/graphalg/playground/}{wildarch.dev/graphalg/playground}.

\begin{figure}[h]
  \centering
  \includegraphics[width=.2\textwidth]{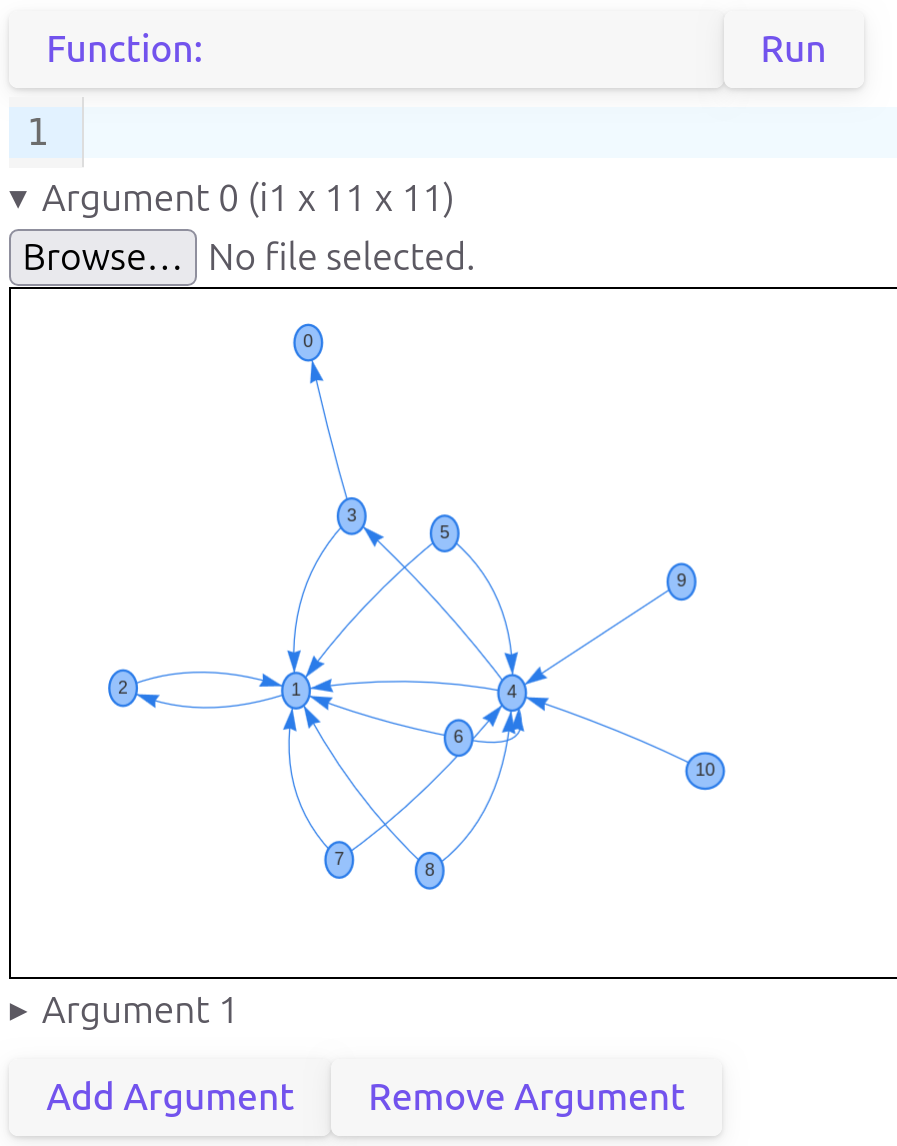}
  \caption{
    \expuser{} uploads a file containing a small graph to the Playground.
  }
  \Description{A file upload dialog in the playground where the user selects a small graph file as input to a GraphAlg program.}
  \label{fig:upload-graph}
\end{figure}

\step{2} \expuser{} uploads a small test graph to the playground (\cref{fig:upload-graph}).

\step{3} \expuser{} writes a sink detection program to verify the test graph has sink nodes; clicking \emph{Run}, the output visualization confirms a sink exists.

\step{4} \expuser{} pastes the original PageRank program into the editor and runs it, noting the score for a well-connected node $n$; summing the computed scores yields less than one, confirming the rank sink issue.

\step{5} \expuser{} adds sink redistribution logic but forgets a required type cast; real-time type checking immediately underlines the error, which they correct.

\step{6} \expuser{} runs the modified program: the score for node $n$ rises, and the scores now sum to one, confirming correct redistribution.

\begin{figure}[h]
  \centering
  \includegraphics[width=.35\textwidth]{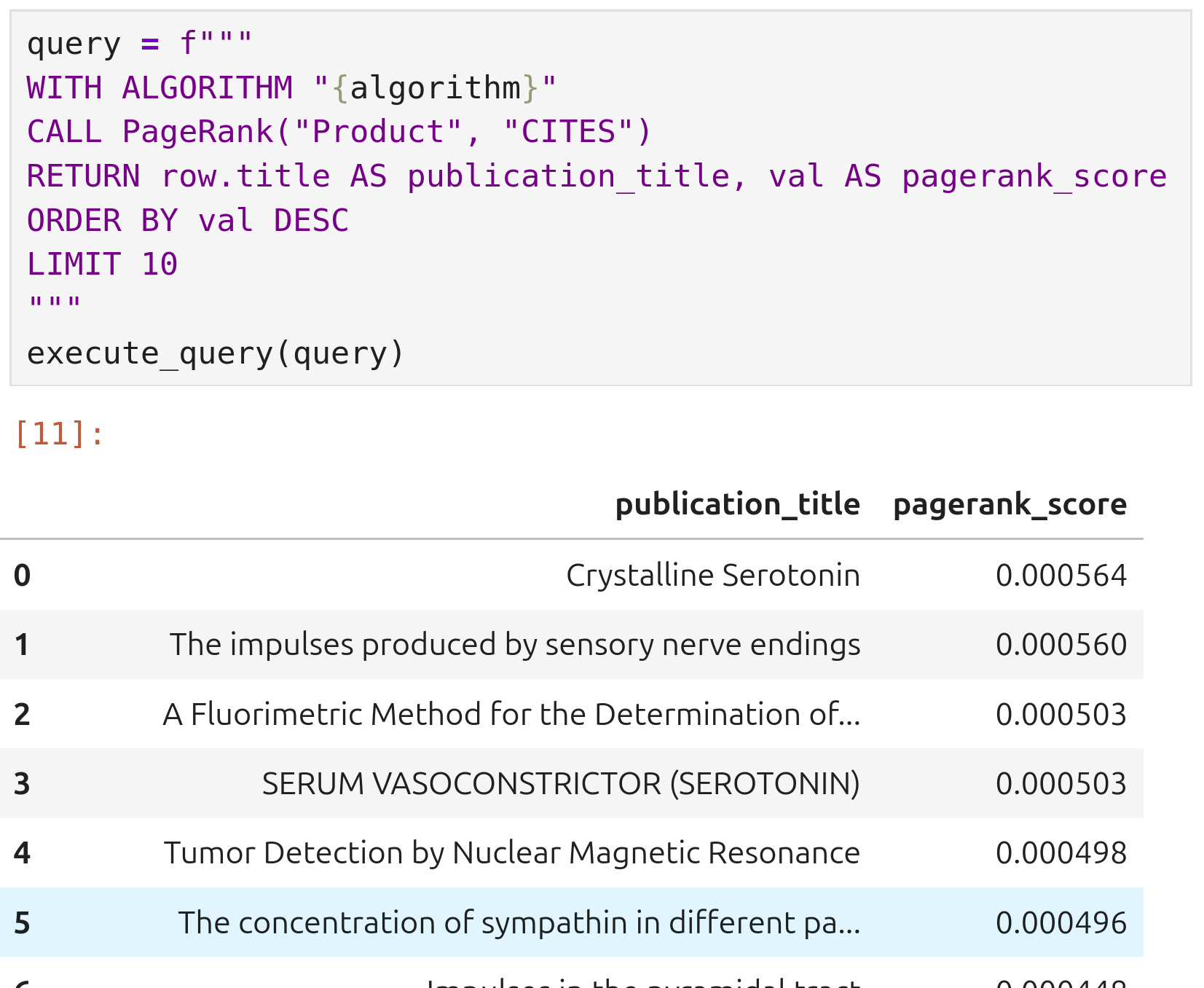}
  \caption{
    \expuser{} executes the PageRank algorithm on the large knowledge graph by sending a Cypher query embedding the algorithm to AvantGraph.
  }
  \Description{A Jupyter Notebook showing a Cypher query with an embedded GraphAlg PageRank program being executed against AvantGraph on a large citation graph.}
  \label{fig:execute-avantgraph}
\end{figure}

\step{7} Confident the algorithm is correct, \expuser{} updates the Cypher query and runs it on the large citation graph (\cref{fig:execute-avantgraph}).

\subsection{Learning \graphalg{} with the Tutorial}
\begin{figure}[h]
  \centering
  \includegraphics[width=.6\linewidth]{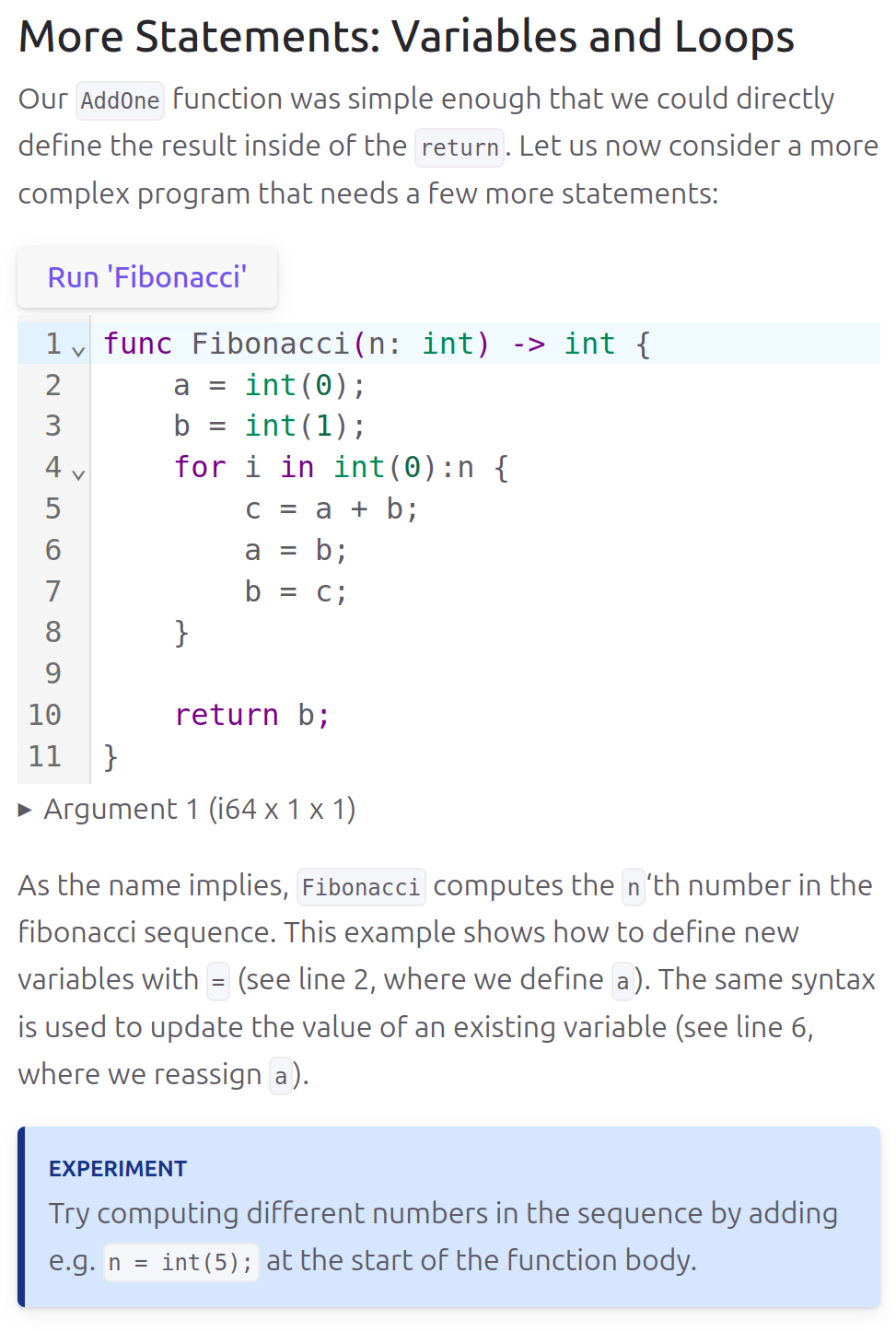}
  \caption{A section of the \graphalg{} tutorial with an interactive code snippet and a suggested experiment.}
  \label{fig:tutorial}
  \Description{A code snippet with a Fibonacci algorithm. Below it is an encouragement to modify the program to compute different values in the Fibonacci sequence}.
\end{figure}

\newuser{} wants to learn \graphalg{} and navigates to the tutorial at \href{https://wildarch.dev/graphalg/tutorial/}{wildarch.dev/graphalg/tutorial}.
They work through the twelve-part tutorial, interacting with code snippets accompanying each concept.
For example, when learning about the Fibonacci sequence (\cref{fig:tutorial}), the tutorial suggests modifying the program. \newuser{} experiments by changing the loop bound and observing how results change.
When they make a mistake, the editor immediately highlights the error.
\emph{Note: Attendees can try the first parts during the session, and complete it later on their own device.}


\begin{acks}
  This work has received funding from the European Union's Horizon Europe framework programme under grant agreement No. 101058573 as part of the SciLake project.
\end{acks}

\bibliographystyle{ACM-Reference-Format}
\bibliography{zotero}


\begin{thebibliography}{22}


\ifx \showCODEN    \undefined \def \showCODEN     #1{\unskip}     \fi
\ifx \showDOI      \undefined \def \showDOI       #1{#1}\fi
\ifx \showISBNx    \undefined \def \showISBNx     #1{\unskip}     \fi
\ifx \showISBNxiii \undefined \def \showISBNxiii  #1{\unskip}     \fi
\ifx \showISSN     \undefined \def \showISSN      #1{\unskip}     \fi
\ifx \showLCCN     \undefined \def \showLCCN      #1{\unskip}     \fi
\ifx \shownote     \undefined \def \shownote      #1{#1}          \fi
\ifx \showarticletitle \undefined \def \showarticletitle #1{#1}   \fi
\ifx \showURL      \undefined \def \showURL       {\relax}        \fi
\providecommand\bibfield[2]{#2}
\providecommand\bibinfo[2]{#2}
\providecommand\natexlab[1]{#1}
\providecommand\showeprint[2][]{arXiv:#2}

\bibitem[\protect\citeauthoryear{Brijder, Geerts, Bussche, and Weerwag}{Brijder
  et~al\mbox{.}}{2019}]%
        {brijder_expressive_2019}
\bibfield{author}{\bibinfo{person}{Robert Brijder}, \bibinfo{person}{Floris
  Geerts}, \bibinfo{person}{Jan Van~Den Bussche}, {and} \bibinfo{person}{Timmy
  Weerwag}.} \bibinfo{year}{2019}\natexlab{}.
\newblock \showarticletitle{On the {Expressive} {Power} of {Query} {Languages}
  for {Matrices}}.
\newblock \bibinfo{journal}{\emph{ACM Trans. Database Syst.}}
  \bibinfo{volume}{44}, \bibinfo{number}{4} (\bibinfo{date}{Oct.}
  \bibinfo{year}{2019}), \bibinfo{pages}{15:1--15:31}.
\newblock
\showISSN{0362-5915}
\urldef\tempurl%
\url{https://doi.org/10.1145/3331445}
\showDOI{\tempurl}


\bibitem[\protect\citeauthoryear{de~Graaf}{de~Graaf}{2026}]%
        {de_graaf_wildarchgraphalg_2026}
\bibfield{author}{\bibinfo{person}{Daan de Graaf}.}
  \bibinfo{year}{2026}\natexlab{}.
\newblock \bibinfo{title}{wildarch/graphalg}.
\newblock
\newblock
\urldef\tempurl%
\url{https://github.com/wildarch/graphalg}
\showURL{%
\tempurl}
\newblock
\shownote{original-date: 2025-10-09T14:03:07Z.}


\bibitem[\protect\citeauthoryear{de~Graaf, Brijder, Chakraborty, Fletcher,
  van~de Wall, and Yakovets}{de~Graaf et~al\mbox{.}}{2026}]%
        {de_graaf_algorithm_2026}
\bibfield{author}{\bibinfo{person}{Daan de Graaf}, \bibinfo{person}{Robert
  Brijder}, \bibinfo{person}{Soham Chakraborty}, \bibinfo{person}{George
  Fletcher}, \bibinfo{person}{Bram van~de Wall}, {and} \bibinfo{person}{Nikolay
  Yakovets}.} \bibinfo{year}{2026}\natexlab{}.
\newblock \bibinfo{title}{Algorithm {Support} for {Graph} {Databases}, {Done}
  {Right}}.
\newblock
\newblock
\urldef\tempurl%
\url{https://doi.org/10.48550/arXiv.2601.06705}
\showDOI{\tempurl}
\newblock
\shownote{arXiv:2601.06705 [cs].}


\bibitem[\protect\citeauthoryear{Francis, Green, Guagliardo, Libkin, Lindaaker,
  Marsault, Plantikow, Rydberg, Selmer, and Taylor}{Francis
  et~al\mbox{.}}{2018}]%
        {francis_cypher_2018}
\bibfield{author}{\bibinfo{person}{Nadime Francis}, \bibinfo{person}{Alastair
  Green}, \bibinfo{person}{Paolo Guagliardo}, \bibinfo{person}{Leonid Libkin},
  \bibinfo{person}{Tobias Lindaaker}, \bibinfo{person}{Victor Marsault},
  \bibinfo{person}{Stefan Plantikow}, \bibinfo{person}{Mats Rydberg},
  \bibinfo{person}{Petra Selmer}, {and} \bibinfo{person}{Andrés Taylor}.}
  \bibinfo{year}{2018}\natexlab{}.
\newblock \showarticletitle{Cypher: {An} {Evolving} {Query} {Language} for
  {Property} {Graphs}}. In \bibinfo{booktitle}{\emph{Proceedings of the 2018
  {International} {Conference} on {Management} of {Data}}}
  \emph{(\bibinfo{series}{{SIGMOD} '18})}. \bibinfo{publisher}{Association for
  Computing Machinery}, \bibinfo{address}{New York, NY, USA},
  \bibinfo{pages}{1433--1445}.
\newblock
\showISBNx{978-1-4503-4703-7}
\urldef\tempurl%
\url{https://doi.org/10.1145/3183713.3190657}
\showDOI{\tempurl}


\bibitem[\protect\citeauthoryear{{Google Inc.}}{{Google Inc.}}{[n.d.]}]%
        {google_inc_go_nodate}
\bibfield{author}{\bibinfo{person}{{Google Inc.}}}
  \bibinfo{year}{[n.d.]}\natexlab{}.
\newblock \bibinfo{title}{Go {Playground} - {The} {Go} {Programming}
  {Language}}.
\newblock
\newblock
\urldef\tempurl%
\url{https://go.dev/play/}
\showURL{%
\tempurl}


\bibitem[\protect\citeauthoryear{Gupta and Ramachandra}{Gupta and
  Ramachandra}{2021}]%
        {gupta_procedural_2021}
\bibfield{author}{\bibinfo{person}{Surabhi Gupta} {and}
  \bibinfo{person}{Karthik Ramachandra}.} \bibinfo{year}{2021}\natexlab{}.
\newblock \showarticletitle{Procedural extensions of {SQL}: understanding their
  usage in the wild}.
\newblock \bibinfo{journal}{\emph{Proceedings of the VLDB Endowment}}
  \bibinfo{volume}{14}, \bibinfo{number}{8} (\bibinfo{date}{April}
  \bibinfo{year}{2021}), \bibinfo{pages}{1378--1391}.
\newblock
\showISSN{2150-8097}
\urldef\tempurl%
\url{https://doi.org/10.14778/3457390.3457402}
\showDOI{\tempurl}


\bibitem[\protect\citeauthoryear{Haas, Rossberg, Schuff, Titzer, Holman,
  Gohman, Wagner, Zakai, and Bastien}{Haas et~al\mbox{.}}{2017}]%
        {haas_bringing_2017}
\bibfield{author}{\bibinfo{person}{Andreas Haas}, \bibinfo{person}{Andreas
  Rossberg}, \bibinfo{person}{Derek~L. Schuff}, \bibinfo{person}{Ben~L.
  Titzer}, \bibinfo{person}{Michael Holman}, \bibinfo{person}{Dan Gohman},
  \bibinfo{person}{Luke Wagner}, \bibinfo{person}{Alon Zakai}, {and}
  \bibinfo{person}{JF Bastien}.} \bibinfo{year}{2017}\natexlab{}.
\newblock \showarticletitle{Bringing the web up to speed with {WebAssembly}}.
  In \bibinfo{booktitle}{\emph{Proceedings of the 38th {ACM} {SIGPLAN}
  {Conference} on {Programming} {Language} {Design} and {Implementation}}}
  \emph{(\bibinfo{series}{{PLDI} 2017})}. \bibinfo{publisher}{Association for
  Computing Machinery}, \bibinfo{address}{New York, NY, USA},
  \bibinfo{pages}{185--200}.
\newblock
\showISBNx{978-1-4503-4988-8}
\urldef\tempurl%
\url{https://doi.org/10.1145/3062341.3062363}
\showDOI{\tempurl}


\bibitem[\protect\citeauthoryear{Hirn and Grust}{Hirn and Grust}{2023}]%
        {hirn_fix_2023}
\bibfield{author}{\bibinfo{person}{Denis Hirn} {and} \bibinfo{person}{Torsten
  Grust}.} \bibinfo{year}{2023}\natexlab{}.
\newblock \showarticletitle{A {Fix} for the {Fixation} on {Fixpoints}}. In
  \bibinfo{booktitle}{\emph{Proceedings of the 13th {Conference} on
  {Innovative} {Data} {Systems} {Research}}}.
\newblock


\bibitem[\protect\citeauthoryear{Hogan, Reutter, and Soto}{Hogan
  et~al\mbox{.}}{2020}]%
        {hogan_-database_2020}
\bibfield{author}{\bibinfo{person}{Aidan Hogan}, \bibinfo{person}{Juan~L.
  Reutter}, {and} \bibinfo{person}{Adrián Soto}.}
  \bibinfo{year}{2020}\natexlab{}.
\newblock \showarticletitle{In-{Database} {Graph} {Analytics} with {Recursive}
  {SPARQL}}. In \bibinfo{booktitle}{\emph{The {Semantic} {Web} – {ISWC}
  2020}}, \bibfield{editor}{\bibinfo{person}{Jeff~Z. Pan},
  \bibinfo{person}{Valentina Tamma}, \bibinfo{person}{Claudia d’Amato},
  \bibinfo{person}{Krzysztof Janowicz}, \bibinfo{person}{Bo~Fu},
  \bibinfo{person}{Axel Polleres}, \bibinfo{person}{Oshani Seneviratne}, {and}
  \bibinfo{person}{Lalana Kagal}} (Eds.). \bibinfo{publisher}{Springer
  International Publishing}, \bibinfo{address}{Cham},
  \bibinfo{pages}{511--528}.
\newblock
\showISBNx{978-3-030-62419-4}
\urldef\tempurl%
\url{https://doi.org/10.1007/978-3-030-62419-4_29}
\showDOI{\tempurl}


\bibitem[\protect\citeauthoryear{Hutchison, Howe, and Suciu}{Hutchison
  et~al\mbox{.}}{2017}]%
        {hutchison_laradb_2017}
\bibfield{author}{\bibinfo{person}{Dylan Hutchison}, \bibinfo{person}{Bill
  Howe}, {and} \bibinfo{person}{Dan Suciu}.} \bibinfo{year}{2017}\natexlab{}.
\newblock \showarticletitle{{LaraDB}: {A} {Minimalist} {Kernel} for {Linear}
  and {Relational} {Algebra} {Computation}}. In
  \bibinfo{booktitle}{\emph{Proceedings of the 4th {ACM} {SIGMOD} {Workshop} on
  {Algorithms} and {Systems} for {MapReduce} and {Beyond}}}.
  \bibinfo{pages}{1--10}.
\newblock
\urldef\tempurl%
\url{https://doi.org/10.1145/3070607.3070608}
\showDOI{\tempurl}
\newblock
\shownote{arXiv:1703.07342 [cs].}


\bibitem[\protect\citeauthoryear{Iosup, Hegeman, Ngai, Heldens, Prat-Pérez,
  Manhardt, Chafi, Capotă, Sundaram, Anderson, Tănase, Xia, Nai, and
  Boncz}{Iosup et~al\mbox{.}}{2016}]%
        {iosup_ldbc_2016}
\bibfield{author}{\bibinfo{person}{Alexandru Iosup}, \bibinfo{person}{Tim
  Hegeman}, \bibinfo{person}{Wing~Lung Ngai}, \bibinfo{person}{Stijn Heldens},
  \bibinfo{person}{Arnau Prat-Pérez}, \bibinfo{person}{Thomas Manhardt},
  \bibinfo{person}{Hassan Chafi}, \bibinfo{person}{Mihai Capotă},
  \bibinfo{person}{Narayanan Sundaram}, \bibinfo{person}{Michael Anderson},
  \bibinfo{person}{Ilie~Gabriel Tănase}, \bibinfo{person}{Yinglong Xia},
  \bibinfo{person}{Lifeng Nai}, {and} \bibinfo{person}{Peter Boncz}.}
  \bibinfo{year}{2016}\natexlab{}.
\newblock \showarticletitle{{LDBC} graphalytics: a benchmark for large-scale
  graph analysis on parallel and distributed platforms}.
\newblock \bibinfo{journal}{\emph{Proceedings of the VLDB Endowment}}
  \bibinfo{volume}{9}, \bibinfo{number}{13} (\bibinfo{date}{Sept.}
  \bibinfo{year}{2016}), \bibinfo{pages}{1317--1328}.
\newblock
\showISSN{2150-8097}
\urldef\tempurl%
\url{https://doi.org/10.14778/3007263.3007270}
\showDOI{\tempurl}


\bibitem[\protect\citeauthoryear{{ISO}}{{ISO}}{2024}]%
        {iso_information_2024}
\bibfield{author}{\bibinfo{person}{{ISO}}.} \bibinfo{year}{2024}\natexlab{}.
\newblock \bibinfo{title}{Information technology — {Database} languages —
  {GQL}}.
\newblock
\newblock
\urldef\tempurl%
\url{https://www.iso.org/standard/76120.html}
\showURL{%
\tempurl}


\bibitem[\protect\citeauthoryear{Kepner and Gilbert}{Kepner and
  Gilbert}{2011}]%
        {kepner_graph_2011}
\bibfield{editor}{\bibinfo{person}{Jeremy Kepner} {and} \bibinfo{person}{John
  Gilbert}} (Eds.). \bibinfo{year}{2011}\natexlab{}.
\newblock \bibinfo{booktitle}{\emph{Graph {Algorithms} in the {Language} of
  {Linear} {Algebra}}}.
\newblock \bibinfo{publisher}{Society for Industrial and Applied Mathematics}.
\newblock
\showISBNx{978-0-89871-990-1}
\urldef\tempurl%
\url{https://doi.org/10.1137/1.9780898719918}
\showDOI{\tempurl}


\bibitem[\protect\citeauthoryear{Ma, Shao, Xiao, Chen, and Wang}{Ma
  et~al\mbox{.}}{2016}]%
        {ma_g-sql_2016}
\bibfield{author}{\bibinfo{person}{Hongbin Ma}, \bibinfo{person}{Bin Shao},
  \bibinfo{person}{Yanghua Xiao}, \bibinfo{person}{Liang~Jeff Chen}, {and}
  \bibinfo{person}{Haixun Wang}.} \bibinfo{year}{2016}\natexlab{}.
\newblock \showarticletitle{G-{SQL}: fast query processing via graph
  exploration}.
\newblock \bibinfo{journal}{\emph{Proceedings of the VLDB Endowment}}
  \bibinfo{volume}{9}, \bibinfo{number}{12} (\bibinfo{date}{Aug.}
  \bibinfo{year}{2016}), \bibinfo{pages}{900--911}.
\newblock
\showISSN{2150-8097}
\urldef\tempurl%
\url{https://doi.org/10.14778/2994509.2994510}
\showDOI{\tempurl}


\bibitem[\protect\citeauthoryear{{Neo4J inc.}}{{Neo4J inc.}}{[n.d.]}]%
        {neo4j_inc_graph_nodate}
\bibfield{author}{\bibinfo{person}{{Neo4J inc.}}}
  \bibinfo{year}{[n.d.]}\natexlab{}.
\newblock \bibinfo{title}{Graph algorithms - {Neo4j} {Graph} {Data} {Science}}.
\newblock
\newblock
\urldef\tempurl%
\url{https://neo4j.com/docs/graph-data-science/2.17/algorithms/}
\showURL{%
\tempurl}


\bibitem[\protect\citeauthoryear{Shaikhha, Suciu, Schleich, and Ngo}{Shaikhha
  et~al\mbox{.}}{2024}]%
        {shaikhha_optimizing_2024}
\bibfield{author}{\bibinfo{person}{Amir Shaikhha}, \bibinfo{person}{Dan Suciu},
  \bibinfo{person}{Maximilian Schleich}, {and} \bibinfo{person}{Hung~Q. Ngo}.}
  \bibinfo{year}{2024}\natexlab{}.
\newblock \showarticletitle{Optimizing {Nested} {Recursive} {Queries}}.
\newblock \bibinfo{journal}{\emph{Proc. ACM Manag. Data}} \bibinfo{volume}{2},
  \bibinfo{number}{1} (\bibinfo{year}{2024}), \bibinfo{pages}{16:1--16:27}.
\newblock
\urldef\tempurl%
\url{https://doi.org/10.1145/3639271}
\showDOI{\tempurl}


\bibitem[\protect\citeauthoryear{Shkapsky, Yang, and Zaniolo}{Shkapsky
  et~al\mbox{.}}{2015}]%
        {shkapsky_optimizing_2015}
\bibfield{author}{\bibinfo{person}{Alexander Shkapsky}, \bibinfo{person}{Mohan
  Yang}, {and} \bibinfo{person}{Carlo Zaniolo}.}
  \bibinfo{year}{2015}\natexlab{}.
\newblock \showarticletitle{Optimizing recursive queries with monotonic
  aggregates in {DeALS}}. In \bibinfo{booktitle}{\emph{2015 {IEEE} 31st
  {International} {Conference} on {Data} {Engineering}}}.
  \bibinfo{pages}{867--878}.
\newblock
\showISSN{2375-026X}
\urldef\tempurl%
\url{https://doi.org/10.1109/ICDE.2015.7113340}
\showDOI{\tempurl}
\newblock
\shownote{ISSN: 2375-026X.}


\bibitem[\protect\citeauthoryear{Sichert and Neumann}{Sichert and
  Neumann}{2022}]%
        {sichert_user-defined_2022}
\bibfield{author}{\bibinfo{person}{Moritz Sichert} {and}
  \bibinfo{person}{Thomas Neumann}.} \bibinfo{year}{2022}\natexlab{}.
\newblock \showarticletitle{User-defined operators: efficiently integrating
  custom algorithms into modern databases}.
\newblock \bibinfo{journal}{\emph{Proceedings of the VLDB Endowment}}
  \bibinfo{volume}{15}, \bibinfo{number}{5} (\bibinfo{date}{Jan.}
  \bibinfo{year}{2022}), \bibinfo{pages}{1119--1131}.
\newblock
\showISSN{2150-8097}
\urldef\tempurl%
\url{https://doi.org/10.14778/3510397.3510408}
\showDOI{\tempurl}


\bibitem[\protect\citeauthoryear{{The DuckDB Developers}}{{The DuckDB
  Developers}}{[n.d.]}]%
        {the_duckdb_developers_duckdb_nodate}
\bibfield{author}{\bibinfo{person}{{The DuckDB Developers}}.}
  \bibinfo{year}{[n.d.]}\natexlab{}.
\newblock \bibinfo{title}{{DuckDB} {Web} {Shell}}.
\newblock
\newblock
\urldef\tempurl%
\url{https://shell.duckdb.org/}
\showURL{%
\tempurl}


\bibitem[\protect\citeauthoryear{{The Rust Developers}}{{The Rust
  Developers}}{[n.d.]}]%
        {the_rust_developers_rust_nodate}
\bibfield{author}{\bibinfo{person}{{The Rust Developers}}.}
  \bibinfo{year}{[n.d.]}\natexlab{}.
\newblock \bibinfo{title}{Rust {Playground}}.
\newblock
\newblock
\urldef\tempurl%
\url{https://play.rust-lang.org/}
\showURL{%
\tempurl}


\bibitem[\protect\citeauthoryear{{The Umbra Developers}}{{The Umbra
  Developers}}{[n.d.]}]%
        {the_umbra_developers_umbra_nodate}
\bibfield{author}{\bibinfo{person}{{The Umbra Developers}}.}
  \bibinfo{year}{[n.d.]}\natexlab{}.
\newblock \bibinfo{title}{Umbra {Online} {Interface}}.
\newblock
\newblock
\urldef\tempurl%
\url{https://umbra-db.com/interface}
\showURL{%
\tempurl}


\bibitem[\protect\citeauthoryear{van Leeuwen, Mulder, van~de Wall, Fletcher,
  and Yakovets}{van Leeuwen et~al\mbox{.}}{2022}]%
        {van_leeuwen_avantgraph_2022}
\bibfield{author}{\bibinfo{person}{Wilco van Leeuwen}, \bibinfo{person}{Thomas
  Mulder}, \bibinfo{person}{Bram van~de Wall}, \bibinfo{person}{George
  Fletcher}, {and} \bibinfo{person}{Nikolay Yakovets}.}
  \bibinfo{year}{2022}\natexlab{}.
\newblock \showarticletitle{{AvantGraph} query processing engine}.
\newblock \bibinfo{journal}{\emph{Proceedings of the VLDB Endowment}}
  \bibinfo{volume}{15}, \bibinfo{number}{12} (\bibinfo{date}{Aug.}
  \bibinfo{year}{2022}), \bibinfo{pages}{3698--3701}.
\newblock
\showISSN{2150-8097}
\urldef\tempurl%
\url{https://doi.org/10.14778/3554821.3554878}
\showDOI{\tempurl}


\end{thebibliography}

\end{document}